\documentclass[multphys,vecphys]{svmult}
\pdfoutput=1
\usepackage{graphicx}        
\usepackage{multicol}        
\usepackage[bottom]{footmisc}

\begin{document}
\title*{Dynamical modeling of the Deep Impact dust ejecta cloud}
\titlerunning{Dynamical modeling of the DI dust ejecta cloud}
\author{
Tanyu Bonev\inst{1}
\thanks{TB is indebted to the organizers 
for supporting his participation
at the conference 
"Deep Impact as a World Observatory Event - 
Synergies in Space, Time, and Wavelength".}\and
Nancy Ageorges\inst{2}\and
Stefano Bagnulo\inst{2}\and
Luis Barrera\inst{3}\and
Hermann B\"{o}hnhardt\inst{4}\and
Olivier Hainaut\inst{2}\and
Emmanuel Jehin\inst{2}\and
Hans-Ullrich K\"{a}ufl\inst{2}\and
Florian Kerber\inst{2}\and
Gaspare LoCurto\inst{2}\and
Jean Manfroid\inst{5}\and
Olivier Marco\inst{2}\and
Eric Pantin\inst{6}\and
Emanuela Pompei\inst{2}\and
Ivo Saviane\inst{2}\and
Fernando Selman\inst{2}\and
Chris Sterken\inst{7}\and
Heike Rauer\inst{8}\and
Gian Paolo Tozzi\inst{9}\and
Michael Weiler\inst{8}
}
\authorrunning{Tanyu Bonev et al.}
\institute{
Institute of Astronomy, Bulgarian Academy of Sciences, Sofia, Bulgaria
\texttt{tbonev@astro.bas.bg}
\and
European Southern Observatory
\and
Universidad Metropolitana de Ciencias de la Educacíon,
Santiago de Chile, Chile
\and
Max-Planck-Institut f\"{u}r Sonnensystemforschung,
Katlenburg-Lindau, Germany
\and
Universit\'{e} de Li\`{e}ge, Belgium
\and
Commissariat Energie Atomique, F-91191 Gif-sur-Yvette, France
\and
Vrije Universiteit Brussel, Belgium
\and
Deutsches Zentrum f\"{u}r Luft und Raumfahrt, Germany
\and
Istituto Nazionale di Astrofisica (INAF) – Osservatorio di Arcetri, Italy
}
%
%
\maketitle
%
%

\setcounter{footnote}{0}
\begin{abstract}
The collision of Deep Impact with comet 9P/Tempel 1 generated a bright cloud
of dust which dissipated during several days after the impact.  The
brightness variations of this cloud and the changes of its position and
shape are governed by the physical properties of the dust grains. We use a
Monte Carlo model to describe the evolution of the post-impact dust plume.
The results of our dynamical simulations are compared to the data obtained
with FORS2\footnote{FORS stands for \underline{FO}cal \underline{R}educer and
low dispersion
\underline{S}pectrograph for the \underline{V}ery \underline{L}arge
\underline{T}elescope (VLT) of the \underline{E}uropean \underline{S}outhern
\underline{O}bservatory (ESO).} to derive the particle size distribution and the total amount
of material contained in the dust ejecta cloud.
\end{abstract}
%
%
\section{Introduction}
\label{sec:1}
Dynamical modeling of the dust coma is often used to
constrain  properties of the dust grains released from a cometary
nucleus. An excellent summary of these efforts is given by \cite{Fulle:04}. 
In his review M. Fulle mentions that, in order to reach a better fit between
model and observation, modelers are often pressed to make the
assumption of time-independent particle size distribution (PSD). 
Observations of
the Deep Impact (DI) dust ejecta cloud represent a special case in this
respect. The time of the impact is exactly known (UT 5:52 on July 4, 2005)
and the time interval 
in which particles excavated by the impact leave the 
circumnuclear region is relatively short 
(with some exceptions which are discussed below).  
Thus, time-dependence of the PSD could be excluded to the first approximation in
model calculations (valid only if no further processing of the particles
takes place later). 
The DI ejecta cloud was observed from the DI spacecraft \cite{AHearn:05} 
and from a great number of ground based observatories \cite{Meech:05}.
We present observations of comet 9P/Tempel 1 (hereafter 9P) obtained
with FORS2 at the VLT of ESO in Paranal. 
We use a Monte Carlo model to describe the DI ejecta cloud observed
during the 4 days after the impact. Inversion of our model allows to derive
the PSD and, 
under appropriate assumptions, to make an estimation of the total mass of dust released by the impact
and ejected with velocities higher than the escape velocity of 9P.
\begin{figure}
\centering
\includegraphics[height=12cm]
{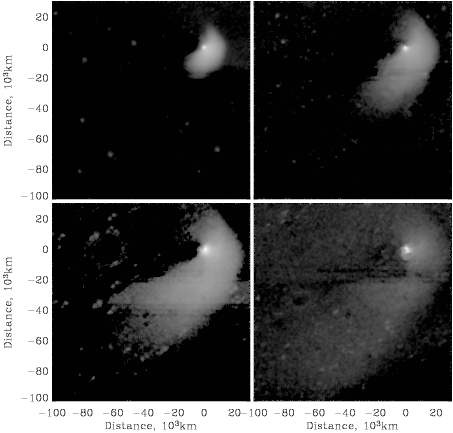}
\caption{Sequence of R-band images representing the dust cloud induced by the
impact. 
From the left upper to the right lower corner the four panels 
show the changing dust distribution during the four post-impact days. 
Each panel contains the difference of the particular R-band image
with a pre-impact image. North is up, East to the left. The projected
direction to the Sun is at 291 degree, counted from North to East}
\label{fig:1}     
\end{figure}
\section{The Observations}
\label{sec:2}
The images used in this analysis were obtained with FORS2\footnote{for
details see http://www.eso.org/instruments/fors1/}
\cite{Appen:98}), mounted at the VLT Antu.
To secure a basis for comparison with the post-impact data several 
images were taken about 6 hours pre-impact, shortly before 
the time when the comet set at Paranal. During the next 4 nights 
images were obtained at 17.8, 42.4, 66.3, 89.4 hours after the impact.
The heliocentric distance of 9P was 1.51 AU, almost constant during the
observing period, as the comet was at perihelion on July 5.35. 
The geocentric distance increased from 0.89 to 0.91 AU, and the pixel scale
changed correspondingly from 162 km/px to 167 km/px.  
The images were calibrated to fluxes and then transformed to Af, the
Albedo-filling factor product \cite{AHearn:84}. 
Figure \ref{fig:1} shows the R-band post-impact images with subtracted
contribution of the pre-impact coma. These images are used in the further analysis. 
\begin{table}
\begin{minipage}[t]{6cm}
The quantity Af is convenient
for comparison with theoretical models as it is directly related to the
total cross-section of the scattering dust particles at any particular
picture element. 
The total scattering cross-section $\times$ Albedo (A $\times$ S)  
of all dust particles produced by the
impact was obtained by integration over the whole area covered by 
the post-impact clouds shown in figure \ref{fig:1}. 
The derived values are presented in table \ref{tab:1}. 
Note the reduction of A $\times$ S with time. 
\end{minipage}
\hfill
\begin{minipage}[t]{5cm}
\centering
\vspace{-4mm}
\caption{Values of the product Albedo $\times$ total scattering cross-section
(A $\times$ S) derived from the post-impact images with subtracted pre-impact coma}
\label{tab:1}   
\begin{tabular}{cc}
\hline\noalign{\smallskip}
Time after   &  A $\times$ S \\
impact, hour &  km$^2$ \\
\noalign{\smallskip}\hline\noalign{\smallskip}
17.8 & 7.6 \\
42.4 & 5.8 \\
66.3 & 5.1 \\
89.4 & 3.2 \\
\noalign{\smallskip}\hline
\end{tabular}
\end{minipage}
\end{table}
\section{The model}
\label{sec:3}
\subsection{Initial conditions}
In the numerical model described below we use particles 
of radii from 0.25 to 250 $\mu$m.
To run the model an initial guess for the velocity dependence on particle
size is needed. The shape of the ejecta cloud (figure \ref{fig:1}) shows
that after the impact the dust expansion is initially at position angle
(P.A.) 240$^\circ$ (the projected direction to the Sun is at P.A. 291$^\circ$).
Later it is deflected by the solar radiation pressure in antisolar direction.
Four days after the impact most of the dust is spread over a large area
in antisolar direction. But even on the fourth day (90 hour
after the impact) dust particles are still found in direction to the Sun. 
We suppose that these are larger particles which are ejected with lower velocities and
which are less influenced by the radiation pressure in comparison
to smaller particles. 
Assuming that the well expressed boundary in direction
to the Sun is the stagnation region of particles ejected with velocity $v$,
and measuring the distance $d$ from the comet to this boundary, we can write
$v\,=\,bt$ and $v^2\,=\,2bd$, where $b$ is the radiation pressure
acceleration. Measured values of $d$, and derived values for the velocity and 
acceleration are given in table \ref{tab:2}. 
The acceleration is used to derive
values for $\beta$, the ratio of gravitational force to the force of radiation
pressure. Finally, from $\beta$, we derive the radii of the particles, $a$, 
using the dependence \cite{Burns:79}: $\beta\,=\,0.585\times10^{-4}Q_{pr}/(\rho
a)$, where $\rho$ is the density of the particles.
The rough estimation of the particles' radii, listed in table \ref{tab:2}, 
was made with an assumed value for the radiation
pressure efficiency, Q$_{pr}$\,=\,1.7, 
with a density, $\rho$ = 1 g\,cm$^{-3}$ 
\begin{table}
\centering
\caption{Apex distance, measured in the four images, and derived values for
the velocity, acceleration, and particle size}
\label{tab:2}
\begin{tabular}{ccccccc}
\hline\noalign{\smallskip}
        UT  & Time after &   Distance     &
{\hspace{1cm}} Terminal {\hspace{1cm}} & Acceleration  & $\beta$ &Particle \\
     Day of &   impact   &                &   
                velocity               &               & values  & radius   \\
July 2005   &   hour     &   10$^3$ km    &    
km/s                                   &    km/s$^2$   &         & $\mu$m    \\
\noalign{\smallskip}\hline\noalign{\smallskip}
   4.972    &   17.8    &   16.5 +/- 2    & 0.51 +/- 0.06  &  8.03e-6  &  3.01 & 0.3 \\
   5.995    &   42.4    &   21.0 +/- 3    & 0.28 +/- 0.04  &  1.81e-6  &  0.68 & 1.5 \\
   6.993    &   66.3    &   25.0 +/- 4    & 0.21 +/- 0.03  &  8.77e-7  &  0.33 & 3.0 \\
   7.955    &   89.4    &   29.0 +/- 5    & 0.18 +/- 0.03  &  5.60e-7  &  0.21 & 4.8 \\
\noalign{\smallskip}\hline
\end{tabular}
\end{table}
Our approach yields the initial velocities of the particles. 
Therefore the values in table \ref{tab:2} are about 2 times larger compared to the mean
velocities given by many other observers (\cite{Meech:05}, \cite{Schleich:06}).  
The derived velocity-size dependence follows a power law with power
index close to -0.4. This dependence resembles the velocities obtained
from theoretical models of the natural gas-dust interaction in the vicinity of a cometary
nucleus (\cite{Gombo:86}, \cite{Crifo:91}), 
as well as the velocities derived from models describing the dust 
coma and tail (\cite{Fulle:04} and references therein).
Initial guess for the location of the impact was found by the constraints
coming from (a)the observed projected expansion direction of the ejected
cloud, (b)the rotation axis orientation (\cite{Belton:06}),
and (c)the latitude of the impact (M. A'Hearn, private communication). 
\subsection{Direct Monte Carlo calculations}
The velocity law and the impact location were
determined more precisely in a process of trial and error. 
We calculated a series of models with values around the initial guess until we
reached a satisfying morphological reproduction of the observed dust
distribution for the four observations. 

In the model used for description of the impact cloud we use
1 million  dust particles which are emitted for a period of 20 minutes 
starting at the moment of the impact. 
These particles are distributed in 100 emission events along  
200 emission directions randomly spaced in a cone with  full opening 
angle of 180 degree. The particles are assumed compact with density of 1
g/cm$^3$ and radii distributed logarithmically in 51 bins, 
in the range from 0.25 to 250 micrometer. After ejection the particles move
along Keplerian orbits under the influence of gravity and radiation
pressure. Their positions are calculated for the times of observation, 
the contributions of the different sizes are weighted with an initial guess
for the PSD and integrated along the line of sight. The modeled brightness,
at position (x,y) is described by:
\begin{equation}
B(x,y) = \sum_{i=0}^{50} K_i \times S_i(x,y)) 
\label{eq:1}
\end{equation}
where S$_i$(x,y) is the scattering area 
produced by the particles  of one particular size, $i$,  
and K$_i$ are the coefficients to be found. 
\subsection{Inversion of the model}
\begin{figure}
\centering
\includegraphics[width=5.5cm, bb=47 134 533 513]
{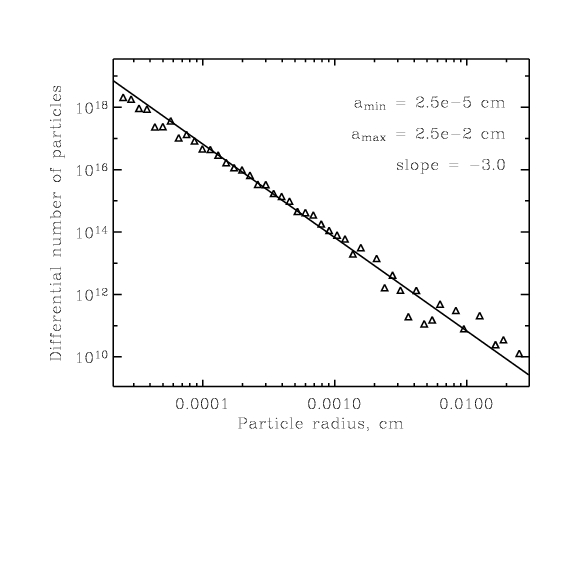}
\hfill
\includegraphics[width=5.5cm, bb=47 134 533 513]
{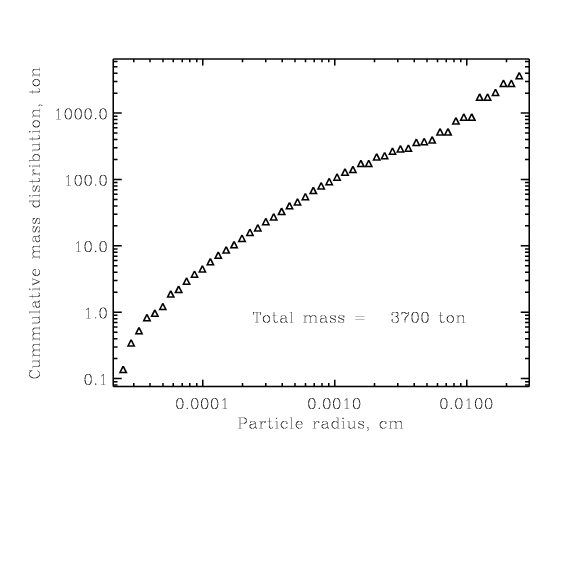}
\caption{The differential particle size distribution (left) and the
cumulative mass distribution (right)
derived from the fit of the model to the observations}
\label{fig:2}     
\end{figure}

The coefficients K$_i$ in equation \ref{eq:1} are derived by 
comparing the modeled brightness $B(x,y)$ to the observed one 
and by minimizing the differences through linear regression. 
Figure \ref{fig:2}
shows the result of the fit to the dust cloud observed 17.8 hours
after the impact. The full line in the left panel 
represents the initial guess for the PSD, a power law with power index -3.0. 
Triangles show the solution of equation \ref{eq:1} converted to the
number of particles of given size.
Particles larger than about 20 micron 
scatter strongly around the mean distribution. 
These large grains are expelled with lower velocities and
their motion is less influenced by the radiation pressure 
in comparison to the smaller particles. 
Therefore large particles of several size bins coexist in a
relatively small region around the nucleus and compete 
in the process of linear regression. 
It seems that we should
simply restrict the upper limit of particle radii to smaller values. Indeed,
this makes the solution more stable, but at the same time it removes the
contribution to the brightness close to the nucleus. In order to reproduce
the enhanced brightness of the ejecta plume observed in this region,
particles of radii as high as 250 $\mu$m are needed. 
Extrapolation of our
initial velocity law shows that these large particles have velocites still above the
escape velocity of 9P, 1.7 m/s \cite{AHearn:05}.  
The right panel in figure \ref{fig:2} 
shows the cumulative mass distribution of the dust. 
We derive a total mass of the dust in the ejecta cloud of 3.7 kiloton.

\begin{figure}
\centering
\includegraphics[height=12cm]
{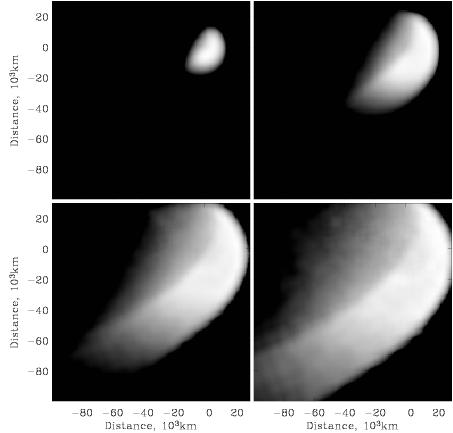}
\caption{Models of the DI dust ejecta cloud 
corresponding to the four observations}
\label{fig:3}     
\end{figure}

The inversion of the model succeeded only in the case of the ejecta cloud
observed 17.8 hours after the impact, which is characterized by the highest
signal-to-noise ratio. The dust distribution in the next images was reproduced
with the parameters found from the fit to the first post-impact observation.  
The four modeled ejecta clouds corresponding to the four 
observations are presented in figure \ref{fig:3}.
\section{Results and discussion}
The PSD derived from the fit to the ejecta cloud observed 17.8 hours after
the impact follows a power law with mean power index -3.0. 
In the detailed model
developed by \cite{Jorda:07} particles of radii $<$ 20 $\mu$m 
are used and a slope of the differential dust size distribution  
-3.2 is found. Light curves of the impact plume were obtained from space  
(\cite{Kuep:05} and from the ground (Pittichova et al. (presentation at
ACM'2005)). These light curves show similar behavior, first a sharp
increase in a time interval dependent on the aperture used,
and second, a gradual decrease of the brightness. 
The decreasing wing of these lightcurves is
well described by subtracting the contribution of small particles
leaving the detector diaphragm with the velocities
used in our model and distributed in accordance with a PSD with slope -3.0.

The velocity distribution of the impact ejecta 
is similar to velocity laws which describe the natural activity of a comet. 
We came to this conclusion empirically, without analyzing possible mechanisms
of particle acceleration related to the impact itself. 
Our conclusion is
based on the particles having velocities greater than 
the escape velocity of 9P.
At the same time our images with removed  pre-impact coma show a brightness
around the nucleus of the comet. Schleicher et al. \cite{Schleich:06}
point to the same feature in their morphological analysis of the
ejecta plume. They explain the enhanced brightness close tot he nucleus 
with the existence of large, heavy particles that
have been ejected with velocities below the escape velocity of the comet.
A future combined analysis of the motion of particles
ejected with velocities above and below the escape limit could be of
interes for understanding the acceleration mechanism of particles produced
by the impact.    

In section \ref{sec:2} we have shown that 
the product scattering area $\times$ Albedo decreases 
with increasing time from the impact. It is 
interesting to extrapolate this trend back in time and to make a comparison 
with data obtained during the first hours after the impact.   
The Rosetta team registered a peak of the light curve 1 hour
after the impact \cite{Kuep:05}. 
From the difference between this value and the
pre-impact level these authors derived a total A $\times$ S
of the newly created dust particles of 33 km$^2$.
This is much greater compared to the value expected from the back
extrapolation of our
measurements, even if we fit them with an exponential law. 
This is an indication that the decrease of A $\times$ S 
should have been faster during the first hours after the impact. Possible
explanation could be the fragmentation of particles with sizes comparable to
the wavelength of observations. Their smaller products will scatter
effectively at shorter wavelengths and, under given conditions, could become
invisible in the R-band. Although this mechanism appears possible further
work is needed to support it with quantitative arguments. 
\section{Conclusions}
We used a dynamical model to describe the dust ejecta created by Deep
Impact. The applied velocity-size dependence was derived empirically. 
Particle with radii in the range 0.25 - 250 $\mu$m were considered. 
We found a best fit to the dust cloud 17.8 hours after the impact with 
differential particle size distribution following a power law with index
-3.0. The total mass of the dust dust ejecta having velocities greater than
the escape velocity is 3700 ton.   
%
%
%

%
%

\end{document}